\documentclass[11pt]{article}   	% use "amsart" instead of "article" for AMSLaTeX format
\usepackage[margin=1in]{geometry}                		% See geometry.pdf to learn the layout options. There are lots.
\usepackage{tipa} 
\usepackage{indentfirst}
\usepackage{graphicx}				% Use pdf, png, jpg, or eps§ with pdflatex; use eps in DVI mode

\usepackage{amssymb,bm,mathrsfs}
\usepackage{amsfonts, amsmath}
\usepackage{subfigure}
\usepackage{float}
\usepackage{booktabs}
\usepackage[flushleft]{threeparttable}
\usepackage{color}
\usepackage{sidecap}
\usepackage{xpatch}
\usepackage[font=small,labelfont=bf]{caption}
\usepackage{mathpazo} % math & rm
\usepackage[scaled]{helvet} % ss
\usepackage{courier} % tt
\usepackage{paralist}

\usepackage[T1]{fontenc}
\usepackage{siunitx} % for the char of temperature
\usepackage{titling}
\normalfont
\usepackage{url}
\usepackage[hyperfootnotes=false]{hyperref}
\usepackage[semicolon,sort&compress]{natbib}
\long\def\symbolfootnote[#1]#2{\begingroup%
\def\thefootnote{\fnsymbol{footnote}}\footnote[#1]{#2}\endgroup}
\def\blfootnote{\xdef\@thefnmark{}\@footnotetext}

\newcommand{\newcite}[1]{\citep{#1}}
\title{\bf{Pre-training of Graph Neural Network for Modeling Effects of Mutations on Protein-Protein Binding Affinity}}
\author{Xianggen Liu$^{1,2,3}$, Yunan Luo$^{1}$, Sen Song$^{2,3}$ and Jian Peng$^{1}$}

\date{}		% Activate to display a given date or no date

\begin{document}
\maketitle
\makeatletter
\xpatchcmd{\paragraph}{3.25ex \@plus1ex \@minus.2ex}{3pt plus 1pt minus 1pt}{\typeout{success!}}{\typeout{failure!}}
\makeatother

\symbolfootnote[0]{\hspace{-0.1in} \footnotesize{$^1$} Department of Computer Science, University of Illinois at Urbana
Champaign, IL, USA.}

\symbolfootnote[0]{\hspace{-0.1in} \footnotesize{$^2$} Laboratory for Brain and Intelligence and Department of Biomedical Engineering, Tsinghua University, Beijing,
China.}
\symbolfootnote[0]{\hspace{-0.1in} \footnotesize{$^3$} Beijing Innovation Center for Future Chip, Tsinghua University, Beijing, China.}

%\symbolfootnote[0]{\hspace{-0.1in} \footnotesize{$^*$} To whom correspondence should be addressed. Email:  \protect\url{songsen@mail.tsinghua.edu.cn}.}

\begin{abstract}
Modeling the effects of mutations on the binding affinity plays a crucial role in protein engineering and drug design. In this study, we develop a novel deep learning based framework, named GraphPPI, to predict the binding affinity changes upon mutations based on the features provided by a graph neural network (GNN). In particular, GraphPPI first employs a well-designed pre-training scheme to enforce the GNN to capture the features that are predictive of the effects of mutations on binding affinity in an unsupervised manner and then integrates these graphical features with gradient-boosting trees to 
perform the prediction. Experiments showed that, without any annotated signals, GraphPPI can capture meaningful patterns of the protein structures. Also, GraphPPI achieved new state-of-the-art performance in predicting the binding affinity changes upon both single- and multi-point mutations on five benchmark datasets. In-depth analyses also showed GraphPPI can accurately estimate the effects of mutations on the binding affinity between SARS-CoV-2 and its neutralizing antibodies. These results have established GraphPPI as a powerful and useful computational tool in the studies of protein design.
\end{abstract}

\section*{Introduction}
Protein-protein interactions (PPIs) play an essential role in various fundamental biological processes. As a representative example, the antibody (Ab) is a central component of the human immune system that interacts with its target antigen to elicit an immune response. This interaction is performed between the complementary
determining regions (CDRs) of the Ab and a specific epitope on the antigen. The ability of Ab to bind to a wide variety of targets in a specific and selective manner has been making antibody therapy wide used for a broad range of diseases including several types of cancer~\newcite{ben2007cancer} and viral infection~\newcite{barouch2013therapeutic}.

Despite the broad application potentials of Ab therapy, it is very challenging to design Abs that have a desired property of PPI. The difficulty lies in two folds. On the one hand, the rational Ab design is usually performed by iteratively mutating residues but the experimental measurement of the PPI change of an Ab mutant is labor-intensive and time-consuming. On the other hand, the number of possible Ab mutants is considerably large, making most of the searching ineffective. Thus, quantitative and fast evaluation of PPI changes upon mutations is crucial for protein engineering and antibody design. In this paper, we mainly focus on the binding affinity of proteins, $\Delta G$, which is one of the most typical properties of PPI.

Traditional methods for modeling the binding affinity changes upon mutation (i.e., $\Delta \Delta G$) can be grouped into three categories: 1) The molecular modeling mechanics, which simulated the free-energy difference between two states of a system based on continuum solvent models~\newcite{lee2006calculation}. 2) Empirical energy-based methods, leveraging classical mechanics or statistical potentials to calculate the free-energy changes, and 3) Machine learning methods that fit the experimental data using the sophisticated engineered features of the changes in structures. The methods in the first category usually provided reliable simulation results but they also required numerous computational resources, which limited their applications. The empirical energy-based methods, exemplified by STATIUM~\newcite{statium}, FoldX~\newcite{schymkowitz2005foldx} and Discovery Studio~\newcite{Discovery}, accelerated the prediction speed but tended to be disturbed by insufficient conformational sampling, especially for mutations in flexible regions. 

As for the machine learning methods, the accumulation of the experimental mutation data has provided an unprecedented opportunity for them to directly model the intrinsic relationship between constructed features of a mutation site and the corresponding binding affinity change. In particular, ~\cite{geng2019isee} proposed a limited number of predictive features from the interface structure, evolution and energy-based features, as the input of a random forest to predict the affinity changes upon mutations. Similarly, MutaBind2~\newcite{zhang2020mutabind2} introduced seven features that described interactions of proteins with the solvent, evolutionary conservation of the site, and thermodynamic stability of the complexes for prediction of the affinity changes upon mutations, which achieved state-of-the-art performance in SKEMPI 2.0 dataset~\newcite{jankauskaite2019skempi}. However, as the features proposed by these machine-learning methods were manually engineered based on the known rules in protein structures, their predictive generalization 
across various protein structures is limited. In addition, these methods relied on the results of the energy optimization algorithms, which was not always reliable due to the stochastic sampling~\newcite{suarez2009challenges,hallen2018osprey}. Here, we are seeking for a method that can generalize well and has less dependency on the performance of third-party optimization algorithms.

%In the fields of natural language process and computer vision, the pre-training of the neural networks provide an attractive solution for generating predictive features for the input data~\newcite{bert,inpainting}. The pre-training scheme can learn intrinsic patterns of data from the numerous unlabeled samples, which could largely benefit the learning of a machine learning model in the fields where data labeling is resource- and time-intensive, such as in the prediction of PPIs.

In this paper, we propose a novel framework, called GraphPPI, to accurately model the effects of mutations on the binding affinity based on the features learned from scratch. In particular, GraphPPI firstly employs a graph neural network (GNN) to produce graphical descriptors and then uses a machine learning algorithm (gradient-boosting trees (GBTs)~\newcite{gbt}) to perform the prediction based on these descriptors. To make the graphical descriptors predictive to the strength of the binding interaction between proteins (i.e., the binding affinity), we introduce a pre-training scheme in which the GNN learns to reconstruct the original structure of a complex given the disturbed one (where the side chains of a residue are randomly rotated).  This reconstruction task requires the GNN to learn the intrinsic patterns underlying the binding interactions between atoms and thus to facilitate the prediction of the GBT.

GraphPPI has the following advantages, compared with traditional methods. 1) GraphPPI is capable to automatically learn meaningful features of the protein structure for the prediction, obviating the need of feature engineering. 2) By training on the large-scale dataset of complexes, the GNN learns the general rules of structures that exist across different complexes, leading to better generalizability of GraphPPI. 

In our experiments, we first investigated what the GNN in GraphPPI learned during the pre-training scheme. We found that, without any annotated labels for learning, GNN can detect the abnormal interaction between atoms and successfully distinguish whether a residue locates in the interface of the complex or not. Second, we evaluated GraphPPI in the prediction of binding affinity changes upon mutations on five benchmark datasets, three for single-point mutations and two for multi-point mutations. GraphPPI obtained the new state-of-the-art performance on these datasets, demonstrating the effectiveness of the graphical features learned by the pre-trained GNN. Besides, we collected several complexes where the newly filtered neutralizing monoclonal antibodies (mAb) bind with the spike glycoprotein protein of SARS-CoV-2. GraphPPI is able to accurately predict the difference in the binding affinity between these complexes, even though the number of mutated residues is larger than that in the training set. Based on one of these mAbs, we also showcased the residues where certain mutations were predicted by GraphPPI to significantly increase the stabilizing effect of the binding with the SARS-CoV-2. These results have demonstrated that our GraphPPI can serve as a powerful tool for the prediction of binding affinity changes upon mutations and have the potential to be applied in a wide range of tasks, such as designing antibodies, finding disease driver mutations and understanding underlying mechanisms of protein biosynthesis.

\section*{Results}

\subsection*{The GraphPPI framework}
\begin{figure}[H]
\centering
\includegraphics[width=0.96\linewidth]
{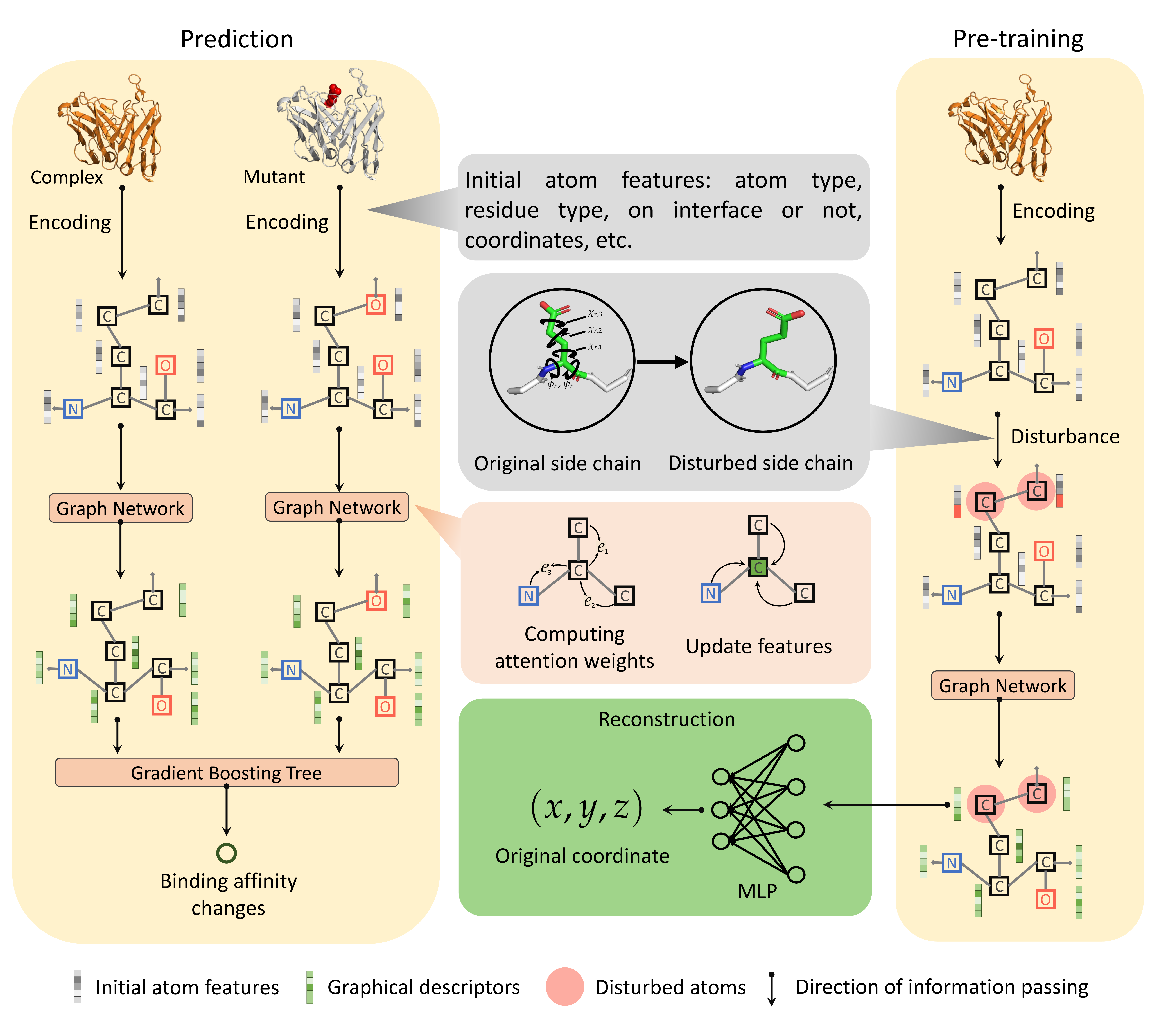}
\caption{The prediction pipeline and the pre-training scheme of the GraphPPI. In the prediction of the GraphPPI, a graph neural network (GNN) produce the graphical descriptors for the given wide-type protein complex and the complex mutant, respectively, and a gradient boosting tree takes these descriptors as the input to predict the corresponding affinity
changes. Before the prediction, a pre-training scheme enforces the GNN to capture the intrinsic patterns underlying the interaction between atoms by reconstructing the original structure of a complex given the disturbed one (where side chains of a residue are randomly rotated).}
\label{fig:framework}
\end{figure}

In this study, we propose a deep learning based framework, name GraphPPI, which models the effects of mutations on the binding affinity based on the features learned by a pre-trained graph neural network (GNN)(Figure~\ref{fig:framework}). More specifically, to achieve both the powerful expressive capacity for graph structures and the robustness of the prediction, the GraphPPI framework accomplishes the prediction by sequentially employing two machine learning components, namely a GNN (excelling in extracting graphical features) and a gradient-boosting tree (GBT, excelling in avoiding overfitting). The GNN integrates the features of neighboring atoms for updating representations of the center atom, providing deep graphical features to represent the complex structure. The GBT takes the graphical features of both the complex and the mutant as inputs and predicts the corresponding binding affinity change. Before the prediction, we need to train these two components to ensure that the GNN can offer useful graphical features for the prediction of the GBT and that the GBT captures the relationship between the inputs (i.e., the graphical features) and the outputs (i.e., the binding affinity changes upon mutations). However, the true features that shape the binding affinity of a complex still remain largely elusive. No supervision signal is given regarding what features that GNN should produce for the input complex.

To address this problem, we proposed a novel pre-training scheme to facilitate the GNN to produce useful graphical features for the input complex. The pre-training in deep learning involves training a model with numerous unlabeled data to obtained deep representations of the input samples, which has been demonstrated to be useful for the various downstream tasks in the fields of natural language processing~\newcite{bert} and computer vision~\newcite{inpainting}.  In the pre-training scheme, the GNN aims to reconstruct the original structure of a complex given the disturbed one where the side chains of a residue are randomly rotated, in which the supervision signals are automatically obtained. In particular, for a protein complex in the pre-training, we disturb the three dimensional coordinates of the side chains of a residue by random rotation. The GNN takes the graphical representations of the disturbed structure as input and learns to reconstruct the original coordinates. 

This well-designed reconstruction task in the pre-training scheme requires GNN to capture intrinsic patterns underlying the interactions between atoms and thus provide predictive graphical features for the downstream task (i.e., prediction of binding affinity changes upon mutations). Considering that most of the previous methods fall into the previously mentioned three categories, GraphPPI is the first attempt (to our best knowledge) to apply a pre-training scheme to extract features for the prediction of binding affinity changes upon mutations.

\subsection*{GraphPPI captures meaningful patterns of the complex structure}
To pre-train the GNN in GraphPPI, we first constructed a large-scale dataset that contains solved structures of 2591 complexes from the PDB-BIND dataset~\newcite{pdb-bind}. The dataset was randomly split into a training set and a development set, with a ratio of 9:1. For each complex, we disturbed the structure by randomly selecting a residue and randomly sampling the angles of its side chains based on the observed distribution~\newcite{shapovalov2011smoothed} (Methods). We repeated the disturbance 10,000 times for each complex, resulting in totally 25,910,000 data points in the dataset. During the training of the GNN, the parameters of GNN that yielded the best performance on the development set were chosen. The
best hyperparameters of GraphPPI were calibrated through a grid search procedure (Supplementary Table~\ref{table:hyper}). The resulting optimal
hyperparameter settings on the development set are listed in Supplementary Table~\ref{table:hyper}. As the binding affinity of the two proteins in the complex is largely determined by the interactions of the atoms in the interface of the two proteins, below, we tried to test 1) whether the pre-trained GNN in GraphPPI can identify the interface region of a complex or not; and 2) whether the pre-trained GNN is sensitive to the abnormal bond length strength of atoms.

\begin{figure}[]
\centering
\includegraphics[width=0.8\linewidth]
{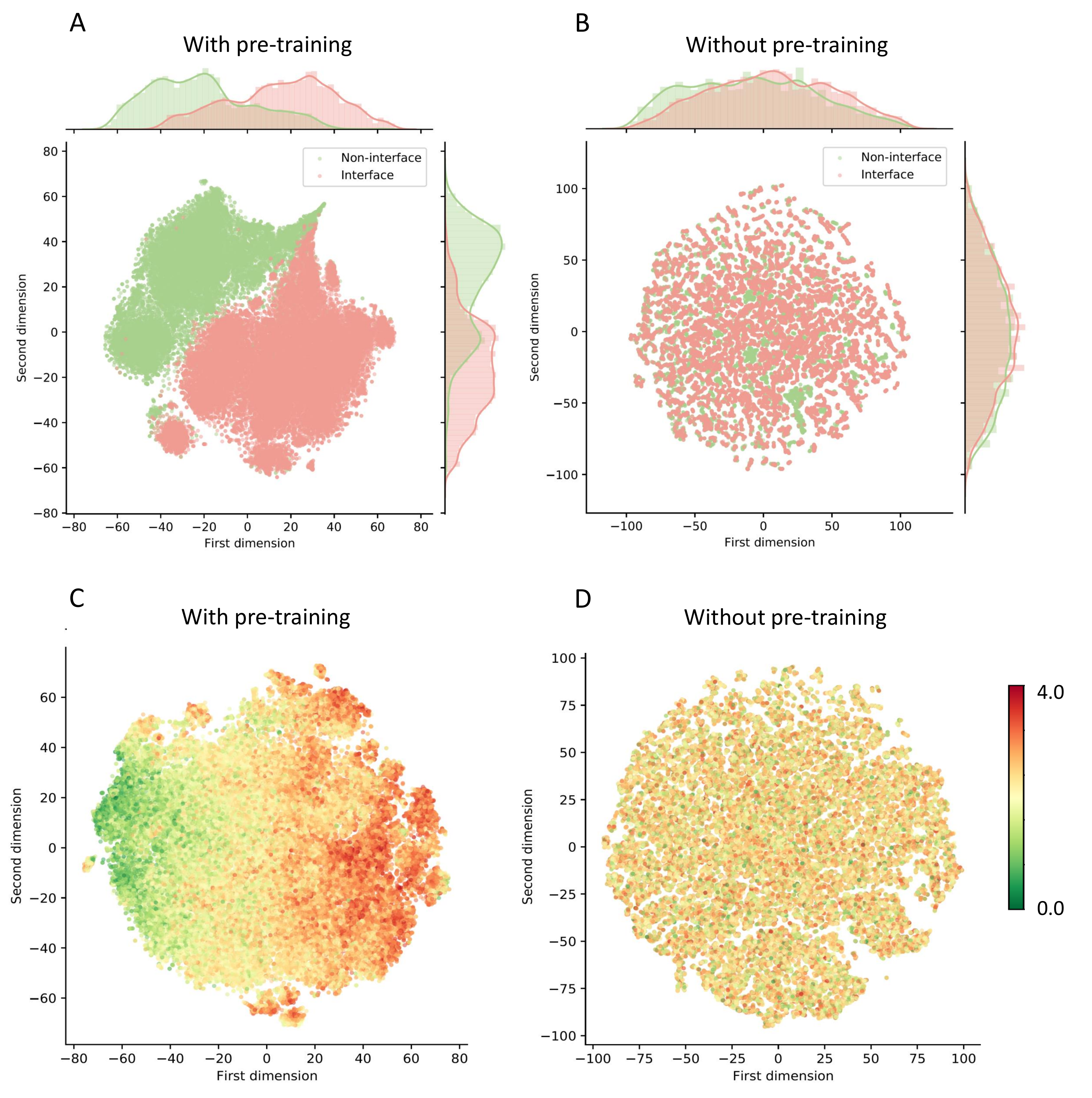}
\caption{Visualization of the graphical descriptors of atoms by t-SNE. A) Distributions of the graphical descriptors of the residues in and not in the interface produced by the pre-trained GNN. B) Distributions of the graphical descriptors of the residues in and not in the interface produced by the GNN without the pre-training scheme. C) Distributions of the graphical descriptors of the atoms with different disturbed distances produced by the pretrained GNN. D) Distributions of the graphical descriptors of the atoms with different disturbed distances produced by the GNN without the pre-training scheme. The disturbed distances were ranging from 0$\si{\angstrom}$ to 4$\si{\angstrom}$.  }
\label{fig:tsne}
\end{figure}

To answer the first question, we fed all the complexes in the dataset into the pre-trained GNN separately and obtained the graphical descriptors of each residue in the interface and those of the non-interface residues. We only keep the same number of the graphical descriptors of the non-interface residues as the interface ones since the non-interface residues are usually more than the interface ones. As $\alpha$-carbon is the central point in the backbone of the amino acid, its graphical descriptors were used to represent the corresponding residue. Then, we employed the t-distributed stochastic neighbor embedding (t-SNE) to visualize the distribution of these graphical descriptors in a low-dimensional space. t-SNE is widely used in machine learning to reduce the feature dimension and preserve the most important two components~\newcite{tsne}. We also performed the same analysis on the GNN model that was not pre-trained for comparison.

Figure~\ref{fig:tsne}A-B shows the distributions of the graphical descriptors of the residues produced by the GNN with or without the pre-training scheme. When we used the GNN that was not pre-trained to generate graphical descriptors, the distribution of the residues in the interface is similar to that of non-interface ones. Expectedly, with the pre-training scheme, the two distributions of the residues in and not in the interface present dramatic distinction in the two dimensional space reduced by t-SNE, indicating our pre-training scheme enables the GNN capable to capture the patterns regarding these key locations (i.e., the interfaces) in a complex. This is largely because the interface and non-interface residues have different solvent accessibilities~\newcite{interfaces}; To accurately reconstruct the original coordinates of the disturbed side chains during pre-training, the GNN have to identify the interface region.

Next, we try to investigate whether the pre-trained GNN can detect the abnormal binding interactions between atoms in a complex. Concretely, for a complex, we randomly selected an atom and disturbed its coordinate within a distance of 4$\si{\angstrom}$. Then we fed the disturbed structure to the GNN and obtained the corresponding graphical descriptors. All the complexes in the pre-training dataset were used and 50 atoms in each complex were randomly selected for this analysis (more atoms did not effect our conclusion). We visualized the distributions of the graphical descriptors of these disturbed atoms produced by the GNN with or without the pre-training scheme in Figure~\ref{fig:tsne}C-D, respectively, where the color of each point indicates the corresponding disturbing distance. We notice that, the graphical descriptors of all the disturbed atoms are scattered uniformly in the space if the GNN were not pre-trained. After the pre-training, the graphical descriptors of the disturbed atoms are arranged orderly in terms of the first dimensions. In other words, the most important component of the graphical descriptors produced by the pre-trained GNN, can clearly indicate the disturbing level of atoms. This comparison demonstrates the pre-trained GNN can detect the abnormal binding interactions between atoms in a complex. This observation is also consistent with the goal of our pre-training scheme, where the GNN learns to reconstruct the abnormal side chains of the complex. Since the mutation in a complex also leads to different conformations (e.g., less stable) from the wide type, the sensitivity of the pre-trained GNN in the
distance of the interactions between atoms may be helpful in the prediction of the binding affinity changes upon mutations. 

\subsection*{GraphPPI advances the state of the art in estimating effects of mutations on binding affinity}
We evaluate GraphPPI on three widely used datasets, namely, the AB-Bind dataset~\newcite{sirin2016ab}, the SKEMPI dataset~\newcite{moal2012skempi} and the SKEMPI 2.0 dataset~\newcite{jankauskaite2019skempi}. Each data point in these datasets comprises the structure of a wide-type complex, the residues to be mutated, the new residue type and the corresponding binding affinity changes. Based on the mutation information, we used the ``buildmodel'' function in FoldX~\newcite{schymkowitz2005foldx} to build the 3D structure of the mutated complex, and transformed the wide-type complex and mutated one the trained GNN to obtain meaningful descriptors for the fitting and prediction of the GBT in GraphPPI. The binding affinity is measured by the binding free energy (Methods), serving as the training labels of the GBT in GraphPPI. 

As the difficulty varies between the prediction tasks in the settings of single-point and multi-point mutations, we assessed the prediction ability of GraphPPI in these two settings separately. The methods that obtained the best results in literature on these two settings are TopGBT~\newcite{wang2020topology} and MutaBind2~\newcite{zhang2020mutabind2}, respectively, which were used as the main baselines in the comparison with GraphPPI. As for the evaluation of the model performance, we mainly considered two metrics, including Pearson correlation coefficient ($R_p$) and root mean square error (RMSE). 

\textbf{Performance on AB-Bind dataset.}
The data points in the AB-Bind dataset are derived from studies of 32 antibody-antigen complexes, each comprising 7 to 246 variants (including both single- and multi-point mutations). The AB-Bind dataset includes 1,101 mutational data points with experimentally measured binding affinities, also denoted by S1101 set. The data come from complexes with crystal structures either of the parent complex or of a homologous complex with high sequence identity. We also followed~\cite{wang2020topology}
to built a sub-dataset that only considers single mutations in the AB-Bind dataset, called the S645 set.

\begin{figure}[]
\centering
\includegraphics[width=0.9\linewidth]
{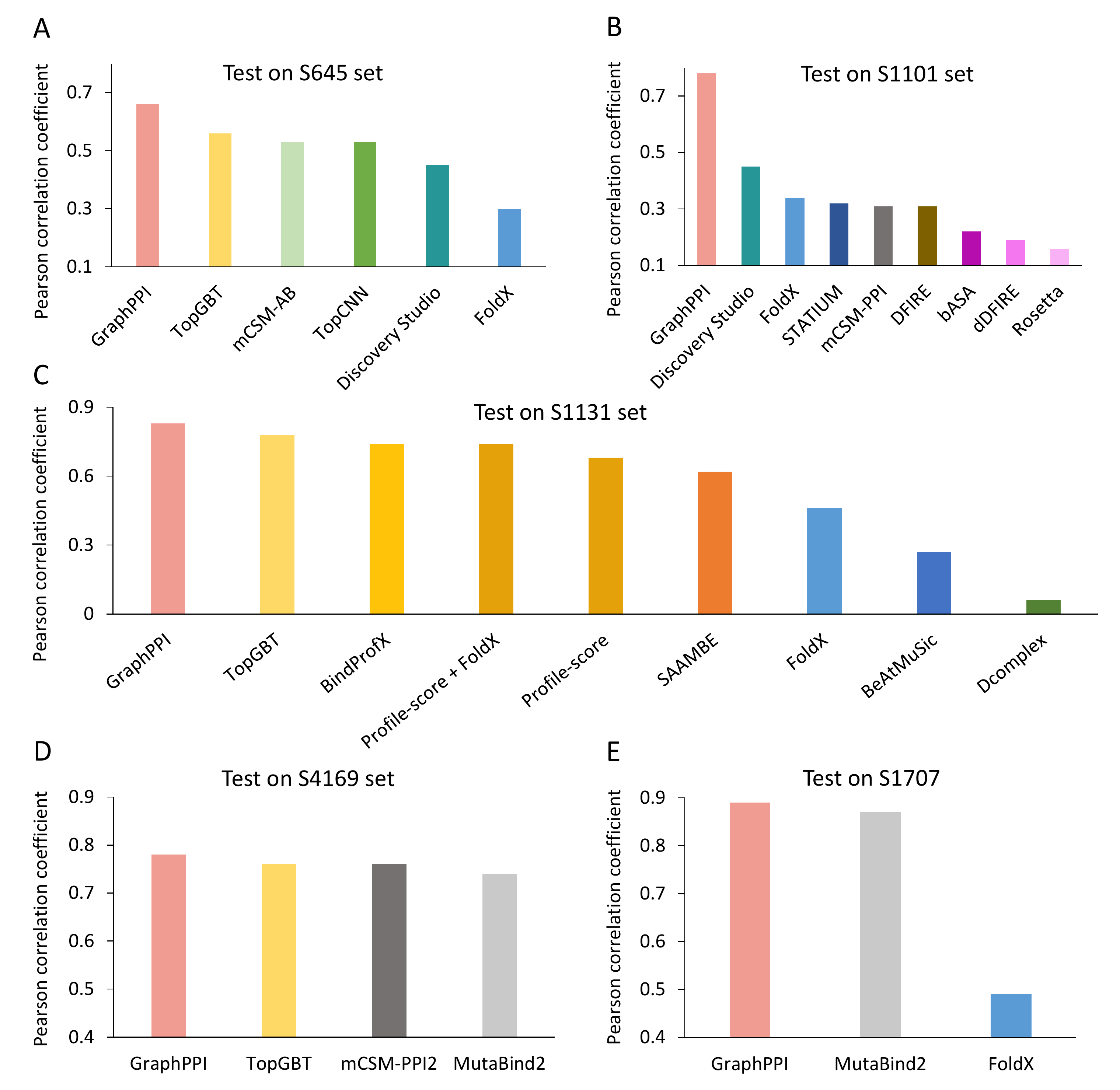}
\caption{Comparison of prediction performance of GraphPPI with that of different baseline methods in terms of Pearson correlation coefficient on the five benchmark datasets.  A) Prediction performance on the S645 set in the AB-BIND dataset, where single residue was mutated in each data point. B) Prediction performance on the S1101 set, containing both single-point and multi-point mutations.  C) Prediction performance on the S1131 set in the SKEMPI dataset, where single residue was mutated in each data point. D) Prediction performance on the S4169 set in the SKEMPI 2.0 dataset, where single residues was mutated in each data point. E) Prediction performance on the S1707 set in the SKEMPI 2.0 dataset, where multiple residues were mutated in each complex. The bar plots shown in (A-D) represent the performance of the ten-fold cross validation test. To have a fair comparison with the MutaBind2 and FoldX, all the results in (E) were obtained by the two-fold cross validation test. All the results except for GraphPPI and TopGBT were quoted from~\cite{wang2020topology}, \cite{sirin2016ab} and~\cite{zhang2020mutabind2}. The performance of TopGBT was obtained by running its published source code.}
\label{fig:perf1}
\end{figure}

The prediction performance of individual methods on AB-Bind S645 is shown in Figure~\ref{fig:perf1}A. We observed that the machine learning methods  (e.g., TopGBT and GraphPPI) obtained better results than the non-machine learning one (i.e., FoldX), which is largely because the non-machine learning method directly makes a prediction based on the optimized results but they do not learn from labeled data. In addition,  our model also behaved significantly better than all the other machine learning models, including the previous state-of-the-art prediction model TopGBT (P<0.01 by paired sample t-test). TopGBT used topology-based features as the representation of the complex, which were originally not designated for representing the patterns of interactions between atoms, limiting their predictive power for binding affinity changes upon mutations. By contrast, the pre-training scheme in GraphPPI was built to explicitly learn the binding interactions of atoms and thus leading into better prediction results.

Further, we use all the data in the S1101 set to evaluate GraphPPI in predicting affinity changes upon both single and multi-point mutations in the ten-fold cross-validation test. Previously, a number of previous methods have been tested on this dataset, such as Discovery Studio~\newcite{Discovery}, STATIUM~\newcite{statium}, mCSM-PPI~\newcite{mcsm-ppi}, DFIRE~\newcite{dfire}, dDFIRE\newcite{dDFIRE}, and Rosetta~\newcite{rosetta}. However, the multi-point mutation involves more complicated molecular dynamics and its impact is not the simple summation of the impact of each mutation, making it more difficult in predicting the corresponding effects. Also, some of the baseline methods do not directly generalize from single-point mutations to multi-point mutations, such as TopGBT and mCSM-AB~\newcite{pires2016mcsm}. Despite these problems, we observe that GraphPPI still yields significantly better performance than all other the baselines, leading to an improvement of 73\% over Discovery Studio~\newcite{Discovery} in terms of $R_P$ (Figure~\ref{fig:perf1}B). Considering that GraphPPI adopts the multi-head attention mechanism, different types of effects induced by individual mutations can be reflected by different dimensions of the graphical descriptors. This distributed fashion of the features better represents the structure changes of multi-point mutations, resulting in the superiority of GraphPPI in prediction performance in this setting. 

\textbf{Performance on SKEMPI dataset.}
The SKEMPI dataset is a database of 3047 binding free energy changes upon mutation assembled from the scientific literature, for protein-protein heterodimeric complexes with experimentally determined structures~\newcite{moal2012skempi}. Subsequently, \cite{xiong2017bindprofx} filtered a subset of 1,131 non-redundant
interface single-point mutations from the original SKEMPI dataset, denoted as S1131 set. A number of previous methods have been tested on this dataset, such as BindProfX~\newcite{xiong2017bindprofx}, Profile-score~\newcite{lensink2013docking}, BeAtMuSiC~\newcite{lensink2013docking}, SAMMBE~\newcite{petukh2016saambe}, Dcomplex~\newcite{liu2004physical}, and TopGBT~\newcite{wang2020topology}. Their prediction performance is shown in Figure~\ref{fig:perf1}C.  Through this comparison, we observed that the methods that directly model the molecular mechanics, such as Dcomplex and FoldX, usually yield poor results. Benefited from the larger number of training data points than AB-Bind s645, the machine learning approaches obtain high correlations between the predicted affinity changes upon mutations and the experimental ones. Also, GraphPPI achieves new state-of-the-art performance on the S1131 set, significantly better than others (P<0.05 by paired sample t-test).

\textbf{Performance on SKEMPI 2.0 dataset.}
As an updated version of the SKEMPI dataset, the SKEMPI 2.0~\newcite{jankauskaite2019skempi} is composed of 7085 single- or multi-point mutations from 345 complexes. \cite{rodrigues2019mcsm} filtered single-point mutations and selected 4,169 variants from 319 different complexes, called S4169 set. On this dataset, GraphPPI achieved an $R_p$ of 0.78 and RMSE of 1.13 kcal mol$^{-1}$, while TopGBT obtained 0.76 and 1.16 kcal mol$^{-1}$ in terms of $R_p$ and RMSE, respectively (Figure~\ref{fig:perf1}D). As for the prediction task for multi-point mutations,  \cite{zhang2020mutabind2} built a dataset called M1707, which contained 1707 data points that have more than one mutation. As shown in Figure~\ref{fig:perf1}E, GraphPPI also achieves the best performance on the M1707 set, achieving an $R_p$ of 0.89 and RMSE of 1.51 kcal mol$^{-1}$, better than the previously best method MutaBind2 (P<0.05 by paired sample t-test). These results further demonstrate the prediction superiority of GraphPPI on both single- and multi-point mutations.

\subsection*{GraphPPI shows better prediction generalizability across proteins and structures}
The generalizability of a machine learning method is our major concern since it determines how broadly a machine learning model can be applied in the prediction of the PPI between proteins. Therefore, in this section, we further evaluate the generalizability of GraphPPI across different proteins and structures.

\begin{figure}[]
\centering
\includegraphics[width=0.95\linewidth]
{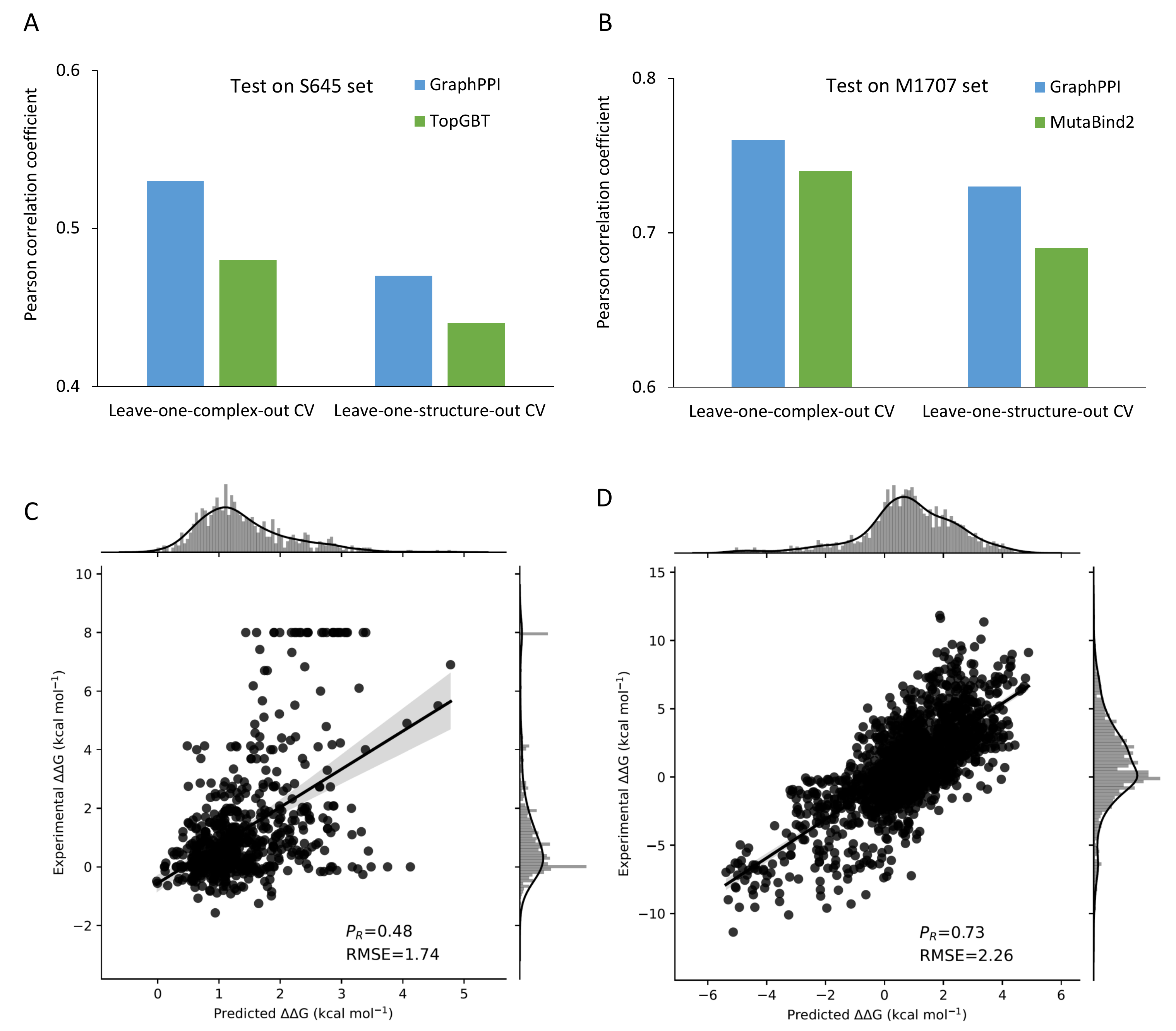}
\caption{Performance of the prediction models in the leave-one-complex-out cross-validation test and the leave-one-structure-out cross-validation test. A) Comparison of the performance of GraphPPI with that of TopGBT on the S645 set. B)  Comparison of the performance of GraphPPI with that of MutaBind2 on the M1707 set. C) The experimental values of the affinity changes and those predicted by GraphPPI in the leave-one-structure-out CV test on the S645 set. D) The experimental values of the affinity changes and those predicted by GraphPPI in the leave-one-structure-out CV test on the M1707 set.}
\label{fig:leave}
\end{figure}

\textbf{Performance of leave-one-complex-out cross-validation.}
In addition to cross-validation tests as we used previously, here we evaluate the models' generalizability by leave-one-complex-out cross-validation (CV) tests on the S645 set (single-point mutation dataset) and M1707 set (multi-point mutation dataset). The leave-one-complex-out CV test involves leaving all the variants of one complex as the test set and using the variants of all the other complexes as the training set. By doing this splitting, the complex in the test set is guaranteed to be not in the training set, which allows us to further estimate the prediction performance of the method on previously unseen data points. In this experiment, we mainly compare GraphPPI with the previous state-of-the-art methods on each benchmark dataset, i.e., TopGBT (on the single-point mutation dataset, Figure~\ref{fig:leave}A) and MutaBind2 (on the single-point mutation dataset, Figure~\ref{fig:leave}B). We found that the prediction performance of all the methods largely decreased compared with that in the ten-fold CV, revealing the more difficulty of the leave-one-complex-out CV test. Also, GraphPPI surpasses TopGBT and MutaBind2 by up to an improvement of 13\% in terms of $R_P$ on the S645 and M1707 sets, respectively (Figure~\ref{fig:leave}A-C). Considering that the features used in MutaBind2 were manually designed, these features were not guaranteed to generalize across different proteins. However, as the features produced by GNN were learned from the large-scale dataset with thousands of types of complexes, the better generalizability across complexes of GraphPPI than MutaBind2 can be expected.

\textbf{Performance of leave-one-structure-out cross-validation.}
As different proteins may share similar structures, a machine learning model has the potentials to overfit the specific structure instead of learning the intrinsic patterns that shape the binding affinity. To further test the generalizability of our model across different structures, we conducted leave-one-structure-out cross-validation (CV) tests on the S645 set and M1707 set. Similar to the leave-one-complex-out CV, the leave-one-structure-out CV test evaluates the prediction model using the variants of each complex as the testing set, but the variants of the other complexes with similar structures were removed from the training set. More specifically, we used TMalign~\newcite{zhang2005tm} software to measure the similarity in structures between two complexes (
Supplementary Figures~\ref{fig:s645-sim} and~\ref{fig:m1707-sim}). As the similarity score over 0.5 in TMalign between two complexes is regarded to share identical parts of structures, we adopted the cutoff of 0.5 to construct the training set (i.e., removing the similar complexes from the training set) for each round of the leave-one-structure-out CV test.

The prediction results of the methods in the leave-one-structure-out CV test are also shown in Figure~\ref{fig:leave}. We found the performance of both GraphPPI and TopGBT has noticeable decreases, compared with the performance in the leave-one-complex-out CV test. This comparison shows that, despite the larger gap between the training and testing data in protein structures, GraphPPI still shows better prediction generalizability  than TopGBT and MutaBind2, again indicating the effectiveness of the pre-training scheme in GraphPPI.

\textbf{Ablation study.}  To investigate the effect of the design choices of GraphPPI, we conducted a series of ablation studies. We first compared the difference of the performance of GraphPPI with and without the pre-training scheme on both ten-fold CV and leave-one-complex-out CV tests. In the GraphPPI without the pre-training scheme, its difference from the original framework is the parameters of the GNN were randomly initialized.  As shown in Supplementary Figure~\ref{fig:ablation},  without a pre-training scheme, GraphPPI shows a significant decrease in prediction performance on the ten-fold CV test, and behaves more poorly in the leave-one-complex-out CV test. These results show that the well-designed pre-training scheme largely improves the predictive power of GraphPPI, especially in the predictive generalizability across different proteins. Next, the ablation study on another key component in GraphPPI, i.e., the GNN model, further confirmed that 
our GNN plays a critical role in learning the interactions between atoms and representing the complex structures than other neural networks (e.g., MLP, Supplementary Figure~\ref{fig:ablation}).

\subsection*{GraphPPI accurately predicts effects of mutations on binding affinities between SARS-CoV-2 and its antibodies}

Severe acute respiratory syndrome coronavirus 2 (SARS-CoV-2) caused an outbreak of pneumonia as a new world pandemic, leading to more than 23 million infection cases and 800 thousand deaths as of August 25, 2020. The spike glycoprotein protein (S) of SARS CoV 2 recognizes and attaches angiotensin-converting enzyme 2 (ACE2) when the viruses infect the human cells. Antibodies that can effectively block SARS-CoV-2 entry into host cells provides a promising therapy for curing the related diseases in the near future. As our framework GraphPPI has shown the powerful predictive capacity in various benchmark datasets, here we wonder whether GraphPPI can capture the effects of mutations of the antibodies on the binding affinity with SARS-CoV-2. 

\begin{figure}[H]
\centering
\includegraphics[width=0.9\linewidth]
{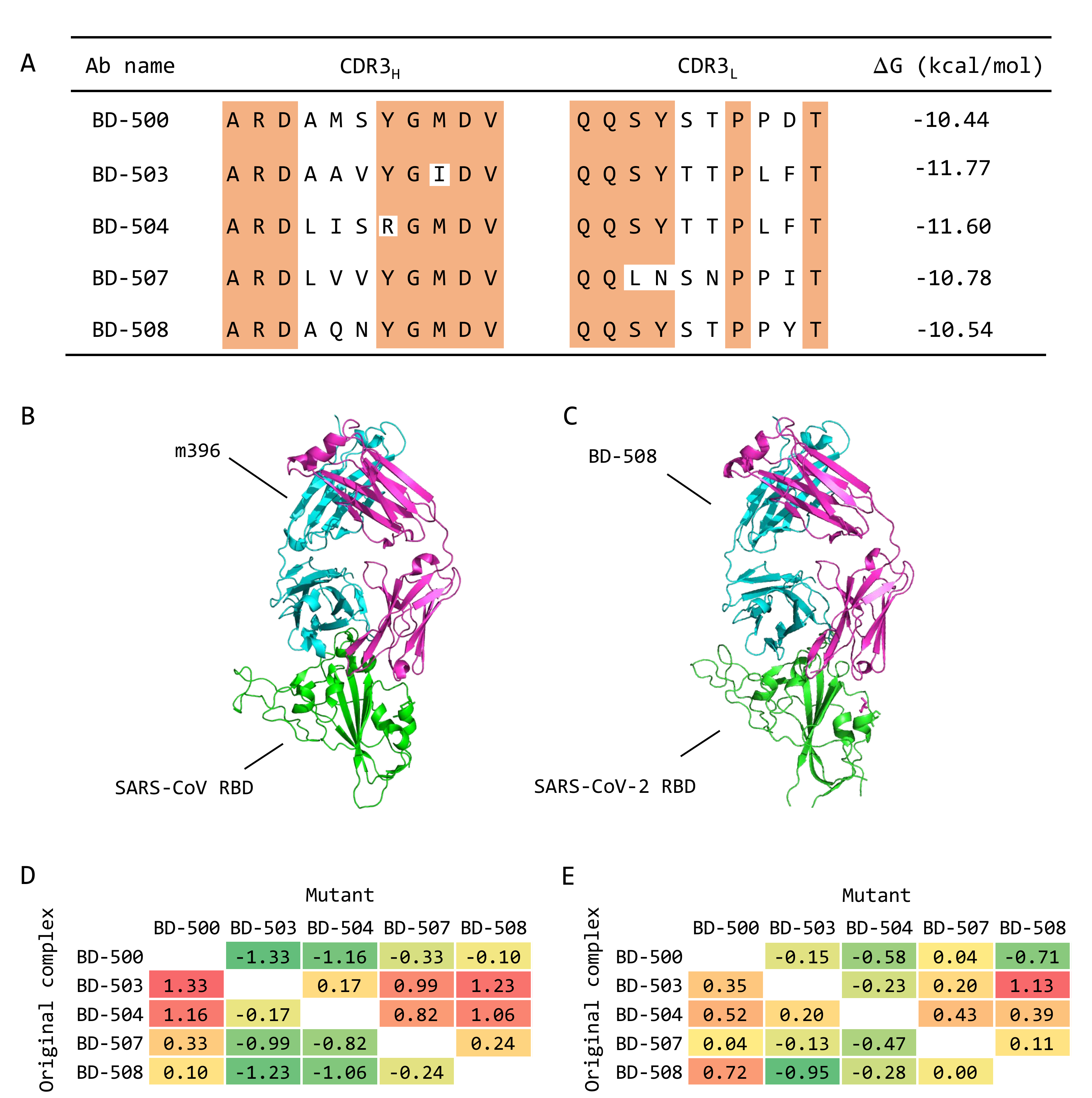}
\caption{The structures of the neutralizing mAbs against SARS-CoV-2 and the comparison of the measured and predicted affinity changes between individual complexes of these mAbs. (A) The CDR3 sequence comparison between SARS-CoV-2 neutralizing mAbs. (B) The solved structure (PDB ID: 2dd8) of the complex composed of m396 and the SARS-CoV-2 RBD. (C) The approximated structure of the complex composed of a mAb (exemplified by BD-508) and the SARS-CoV-2 RBD. (D) The observed binding affinity changes between two arbitrary mAbs. (E) The binding affinity changes between two arbitrary mAbs predicted by GraphPPI.}
\label{fig:mabs}
\end{figure}

To this end, we constructed a small dataset that contains several complexes of the SARS-CoV-2 S protein with different monoclonal antibodies (mAbs). We collected five potent neutralizing mAbs against SARS-CoV-2 from \cite{cao2020potent} (Figure~\ref{fig:mabs}A). These mAbs neutralize the SARS-CoV-2 by binding with the receptor binding domain (RBD) of the S protein with different binding strength. \cite{cao2020potent} measured their binding affinities with SARS-CoV-2 using surface plasmon resonance (SPR) and also showed that these mAbs shared high homology in the CDR3$_H$ sequences with SARS-CoV neutralizing mAb m396 (PDB ID: 2dd8, which neutralizes SARS-CoV by disrupting ACE2/RBD
binding), providing a way to approximate the structures of these five mAbs. Based on the solved three-dimensional (3D) structure of m396, we leveraged the ``buildmodel'' function in FoldX to construct the 3D structures of these mAbs and then used ZDOCK software~\newcite{pierce2014zdock} to predict the orientations of the RBD to these mAbs (Figure~\ref{fig:mabs}C). Finally, we obtained the approximated structures of the five complexes that comprise SARS-CoV-2 RBD with neutralizing 
mAbs. These constructed complexes are relatively reliable since they share similar structures and binding sites with the complex of SARS-CoV bond by m396 (Figure~\ref{fig:mabs}B-C).

As these collected complexes can be regarded as multi-point missense mutants to each other (Figure~\ref{fig:mabs}A), we evaluated the performance of GraphPPI by measuring the difference in the predicted and observed affinity changes between each pair of these complexes. Before testing, GraphPPI was trained on all the data points of the large-scale M1707 dataset.

Figure~\ref{fig:mabs}D shows the measured binding affinity changes between individual pairs of the complexes of mAbs and Figure~\ref{fig:mabs}E shows the corresponding affinity changes predicted by GraphPPI. GraphPPI achieved a strong correlation of 0.71 and an RMSE of 0.67 kcal mol$^{-1}$, suggesting that despite the larger number of mutated points than that in training data, GraphPPI still successfully captures their changes of binding affinity upon mutations to a large extent.

\subsection*{GraphPPI provides biological insights in designing antibody against SARS-CoV-2}

As GraphPPI can accurately predict the affinity changes upon mutations of the mAbs complexes  against SARS-CoV-2, here we try to use GraphPPI to design mAb mutants that can bind with SARS-CoV-2 with better stability (measured by the negative change in binding free energy). As a case study, we performed a one-step design on the base of BD-508, a mAb verified in the plaque reduction neutralization test (PRNT) using authentic SARS-CoV-2 isolated from COVID-19 patients and showed high potencies~\newcite{cao2020potent}. We performed a full computational mutation scanning on the interface of the complex consisting of BD-508 and the SARS-CoV-2 RBD to investigate which mutations tend to yield higher binding affinities. The 34 residues of BD-508 were mutated to all other 19 amino acid types. To reduce the computational complexity of the scanning, we focused on single-point mutations. In total, we thus conducted 646 computational mutations. We collected the data points with single-point mutations from the S4169 set and also included their reversed mutations by setting the corresponding affinity changes to the negative values of its original ones, obtaining a dataset with 8338 variants. We adopted this dataset to train GraphPPI and used the trained GraphPPI to predict the binding affinity changes of the above 646 mutations on the interface of BD-508 separately. 

\begin{figure}[]
\centering
\includegraphics[width=0.85\linewidth]
{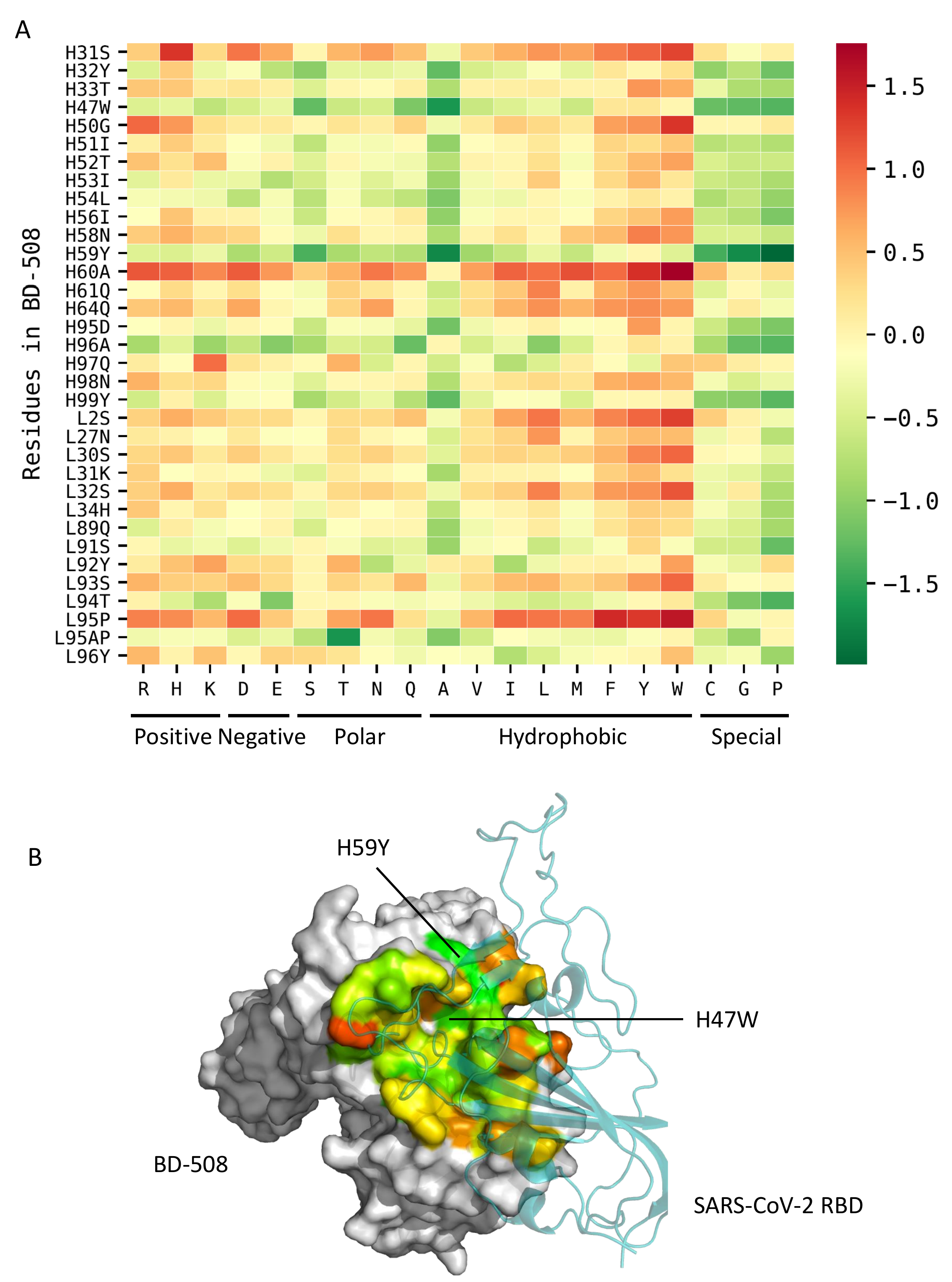}
\caption{The affinity changes of each residue with different mutations in the interface of the complex of BD-508 with the SARS-CoV-2 RBD. (B) The visualization of the average affinity changes of the mutations in each residue in the interface of DB-508.}
\label{fig:scaning}
\end{figure}

The affinity changes of each residue with different mutations were shown in Figure~\ref{fig:scaning}A, where the amino acid types of the mutants were categorized into charged, polar, hydrophobic and special-case groups.
We observe that the affinity changes were highly correlated with the amino acid types and mutating to tryptophan often yields the highest binding affinity changes in this case. In addition, there are three sensitive residues in BD-508 whose mutations could significantly improve the binding affinity, i.e., H47W and H59Y (Figure~\ref{fig:scaning}B). Through this case study and the tests on the mAbs against SARS-CoV-2, we demonstrate that GraphPPI can accurately predict the binding affinity changes upon mutations of mAbs and showcase how to use GraphPPI to improve the binding affinity of mAbs with SARS-CoV-2. We believe our GraphPPI can serve as a useful tool in designing antibodies and other related biological tasks. 

\clearpage
\section*{Methods}
\subsection*{Definition of the task of predicting protein-protein binding affinity changes upon mutation}
Given 3D structure of a protein-protein complex, the residue(s) to be mutated and the new residue type(s), the goal is to estimate the binding free energy changes (i.e., $\Delta \Delta G$) between the original complex with the mutant. 

\begin{equation}
    \Delta \Delta G = \Delta G_{\text{mutant}}-\Delta G_{\text{wide-type}},
\end{equation}
where $\Delta G$ is the binding free energy of a complex.

\subsection*{Detailed implementation of individual modules in GraphPPI}
\subsubsection*{Encoding the graph representations of a given complex}

To build a graph for a given protein complex, we regard atoms as nodes, and their interactions as edges. We only consider four types of atoms, namely C, O, N, S. For each node $k$, we use its attributes as its raw features, such as element type, the residue type, the chain index, the location information (i.e., is on the interface or not), and the three dimensional coordinate (i.e.,$({x}_k,{y}_k,{z}_k)\in \mathbb{R}^3$). All attributes we used and their encoding techniques are specified in Supplementary Table~\ref{tab:features}. We concatenate their encodings into a vector. The vector is $D=30$ dimensional and used as the initial features of the corresponding node. The features of all the nodes are denoted as $\bm A\in\mathbb{R}^{
N\times D}$, where $N$ stands for the number of nodes. As for the edges, if the distance of two nodes is shorter than a threshold (i.e., 3$\si{\angstrom}$), we assume there exists an edge between them. The connected edges on the entire complex are denoted as $\bm E\in\mathbb{R}^{N\times N}$, in which the entries are either one or zero. Therefore, for a given complex, its initial graph representations are $(\bm A, \bm E)$. %The strength of connected edges varies across the different connected atoms. Thus we divide the types of edges into ten categories, each corresponding to a combination of above two atoms. We use $\bm C\in\mathbb{R}^{N\times N}$ to indicate the categories of edges.

\subsubsection*{Generation of deep graphical features by graph neural network (GNN)}
To capture the structure of a protein complex at the atom level, we propose a graph neural network named coordinate-based graph attention network (CGAT, Figure~\ref{fig:framework}). CGAT shares the basic idea of graph attention network (GAT)~\newcite{velivckovic2017graph}: for each node, CGAT uses the representations of the neighboring nodes to update its representation. But different from GAT, CGAT specifically considers the coordinates in the input vectors when controlling the update process. 

Specifically, given the atoms (also called nodes generally) features $\bm A$ and the edges $\bm E$, GAT learns to capture the interaction information between atoms. Formally, GAT  performs self-attention mechanism on the nodes to indicate the importance of node $j$'s features to node $i$, computed by 
\begin{equation}
    \label{eq:score}
    s_{i,j}=\text{LeakyReLU}(\bm u^T[\bm W\bm A_i||\bm W\bm A_j]),
\end{equation}
where $\bm A_i$, the $i$-th vector in $\bm A$, stands for the features of the node $i$. $||$ represents concatenation, and LeakyReLU stands for LeakyReLU nonlinear function~\newcite{he2015delving}.  $\bm W\in\mathbb{R}^{D \times D_g}$ and $\bm u\in\mathbb{R}^{D}$ are learnable weight matrix and vector, respectively. $D_g$ is the hidden size in GAT. In the initial graph representations of the complex, the absolute values of three dimensional coordinates in the node features vary a lot across different complexes, the difference between coordinates is more useful. Thus, CGAT also integrates the difference of the two atoms into the self-attention mechanism at the first transformation layer.
\begin{equation}
    \label{eq:score}
    s_{i,j}=\text{LeakyReLU}(\bm u^T[\bm W\bm A_i||\bm W\bm A_j||\bm W'(\bm A_i - \bm A_j)]),
\end{equation}
where $\bm W' \in\mathbb{R}^{D \times D_g}$ is learnable weight matrix.

To make coefficients easily comparable across the different nodes adjacent to node $i$, we then normalize them across all choices of $j$ using the softmax function:
\begin{equation}
    e_{i,j} = \text{softmax}(s_{i,j}) = \frac{e^{s_{i,j}}}{\sum_{j\in \mathcal{N}_i} e^{s_{i,j}}} ,
\end{equation}
where $\mathcal{N}_i$ stands for the neighbors of node $i$. Once obtained, the normalized attention coefficients together with the corresponding atom features are used to apply weighted summation operation, resulting into the updated representations of node $i$, given by
\begin{equation}
\label{eq:sum}
\bm h_i = \delta (\sum_{j\in \mathcal{N}_i} e_{i,j} \bm W \bm x_j),
\end{equation}
where $\delta(\cdot)$ represents nonlinear function, e.g., ReLU function~\newcite{he2015delving}. The computations of Equations (\ref{eq:score}-\ref{eq:sum}) form a transformation module, called a self-attention layer.

CGAT also employs multi-head attention to stabilize the learning process of self-attention, that is, $K$ independent attention mechanisms execute the transformation of Equation~(\ref{eq:sum}), and then their features are concatenated, resulting in the following feature representations.
\begin{equation}
\bm h_i =\text{M-Attention}(\bm X,\bm x_i)= ||_k^K \delta (\sum_{j\in \mathcal{N}_i} e_{i,j}^k \bm W^k \bm x_j),
\end{equation}
where $e_{i,j}^k$ are normalized attention coefficient computed by the $k$-th attention mechanism, and $W^k$ is the corresponding input linear transformation's weight. Note that, in this setting, the final returned output $h_i$ will consist of $KD_g$ features (rather than $D_g$) for each node.

To extract a deep representation of the product and increase the expression power of the model, we stacked $L$ multi-attention layers.
\begin{equation}
\label{eq:multi-att}
\bm h^{(l+1)}_i = \text{M-Attention}(\bm H^{(l)},\bm h^{(l)}_i), l=1,2,\dots,L,
\end{equation}
where $\bm H^{(l)}$ stands for the  features of all the nodes processed by $l$-th layer of GAT and $\bm h^{(l)}_i$ indicates processed features for node $i$. Based on the node features processed by the last layer, to further enlarge the receptive field of the transformation for each node and encourage larger values in node features, we also employ max-pooling function to gather the information from the neighboring node as part of the final graphical descriptors of nodes.
\begin{align}
\label{eq:max}
g_{i,j} &=  \max_{j\in \mathcal{N}_i}  \bm h_j^{(L)},\\
\bm g_i &= ||_{j=1}^{j=D_g} \ g_{i,j} ||\ \bm h_i^{(L)}.
\end{align}
That is to say, given the initial node features $\bm A$ and the connected edges $\bm E$ of a complex, the GNN outputs graphical descriptors $\bm g_i$ for the atom $i$ in the complex.
\begin{align}
{\bm G} = \{\bm g_i|i=1,2,\dots,N\} = \text{CGAT}({\bm A}, \bm E),
\end{align}
where $\bm G$ is the set of the graphical descriptors of all the atoms.

\subsubsection*{Prediction of binding affinity changes upon mutations by gradient-boosting tree (GBT)}

For the prediction of the binding affinity change $y\in \mathbb{R}$ given the original protein complex $x$ and its mutant $m$, GraphPPI integrates the graphical descriptors in the pre-trained GNN with a gradient-boosting tree (GBT)~\newcite{gbt}. In particular, GraphPPI first leverages the pre-trained CGAT to generate features that are expected to represent the affinity change from the original complex to its mutant. For both original protein $x$ and its mutant $m$, the learned graphical descriptors of each atom at the mutated residues (denoted as $\bm G_{xm}$ and $\bm G_{mm}$, respectively) and the learned graphical descriptors of each atom at the interface residues (denoted as $\bm G_{xi}$ and $\bm G_{mi}$, respectively) are selected, which are given by
\begin{align}
 \bm G_{xm} &= [ \bm g_k], k\in S_{xm},\\
 \bm G_{mm} &= [ \bm g_k], k\in S_{mm},\\ 
 \bm G_{xi} &= [ \bm g_k], k\in S_{xi},\\ 
 \bm G_{mi} &= [ \bm g_k], k\in S_{mi},
\end{align}
where $S_{xm}$ stands for the set of atoms that belongs to the residue to be mutated in the original complex.  $S_{mm}$ stands for the set of atoms that belongs to the residue mutated in the mutant complex. $S_{xi}$ stands for the set of atoms that belongs to the interface residues in the original complex. $S_{mi}$ stands for the set of atoms that belongs to the interface residues in the mutant complex. 

Due to the specific design in extracting graphical descriptors in the CGAT (such as the ReLU function and Equation~(\ref{eq:max})), the larger values of features represent higher importance. Therefore, for each collection of the graphical descriptors (namely $\bm G_{xm},\bm G_{mm},\bm G_{xi},\bm G_{mi}$), we use max-pooling and mean-pooling operations to obtain their mean values and max values at each dimension over the selected atoms, that is, 
\begin{align}
& \bm F_{n} = [\text{maxpooling}(\bm G_{n}), \text{meanpooling}(\bm G_{n})], \\ 
& \text{maxpooling}(\bm G_{n})_i = \max \{\bm g_k, k\in S_n\},\\
& \text{meanpooling}(\bm G_{n})_i = \text{mean} \{\bm g_k, k\in S_n\}, n\in\{xm, mm, xi,mi\},
\end{align}
where $i$ denotes the dimension index of the learned representations.

Finally, gradient boosting tree (GBT) takes these features as input to accomplish the prediction of the binding affinity changes (i.e., $\Delta\Delta G$) upon mutations. 
\begin{align}
\label{eq:gbt}
\Delta\Delta G = \text{GBT}([\bm F_{xm},\bm F_{xm},\bm F_{xi},\bm F_{mi}]).
\end{align}

\subsection*{Detail implementation of the pre-training scheme}
Pre-training strategies have been demonstrated to be powerful in various applications, such as computer vision~\newcite{inpainting} and natural language processing~\newcite{bert}. The pre-training of graph networks also shows significant performance gains in the task of the prediction of small molecular properties~\newcite{Hu2020Strategies}. However, due to the complexity of the dynamics in protein structure, no pre-training scheme has been studied in this field. In this paper, based on the characters of protein conformations, we carefully design a novel pre-training scheme which is specific for the prediction of the affinity changes upon mutation. Generally speaking, in the proposed pre-training scheme of GraphPPI, the GNN (specifically CGAT) aims to reconstruct the original structure of a complex given the disturbed one where side chains of residue are randomly rotated. Below, we will elaborate the disturbance procedure and the reconstruction process.

\textbf{Disturbance.} To produce some meaningful disturbances in the given complex, we propose to rotate the side chains of a randomly selected amino acid. This idea stems from the observation that, for a particular complex, only a few conformations can lead to the lowest free energy. Most of the disturbances will increase the free energy and make the complex less stable. By reconstructing the original conformations, a model is expected to capture the patterns of the biomolecular interactions between atoms and those between residues in the three dimensional space.

Formally, let $r$ be a certain residue and $\phi_r,\psi_r$ be its two dihedral angles near the alpha carbon (Figure~\ref{fig:framework}). The disturbed side chains of residue $r$ are sampled from the distribution of corresponding side-chain conformations. That is,
\begin{align}
\bm \chi_r \sim p(\cdot|\phi_r,\psi_r, r)p(r),
\end{align}
where $p(\cdot|\phi_r,\psi_r, r)$ stands for the distribution of the side-chain conformations of the residue $r$, which is approximated by a protein-dependent side-chain rotamer library proposed by~\cite{shapovalov2011smoothed}. $p(r)$ describes the probability of the residue $r$ being selected during the disturbance. As our downstream task is to model the binding affinity, which is usually characterized by the interface residues of the complex, we set $p(r)$ to be the uniform distribution over the interface residues and the probabilities of other non-interface residues are zero. Note that individual amino acids may have different numbers of side chains. For notational simplicity, we use $\bm \chi_r$ to be the set of the angles of the side chains. Taking the Glutamic acid for example, There are three side chains, that is, $\bm \chi_r =(\chi_{r,1},\chi_{r,2},\chi_{r,3}) $.

Based on the sampled side chains and the coordinates of the backbone of the original residue $r$, we can derive the new coordinates of each atom in residue $r$, which are given by
\begin{align}
(\hat{x}_k,\hat{y}_k,\hat{z}_k)  = \text{Coordinates} (k,\bm \chi_r,r), k\in S(r),
\end{align}
where $\text{Coordinates}((k,\bm \chi_r,r))$ stands for the function that yields the coordinates of atom $k$ based on the side chains of residue $r$. $S(r)$ stands for the set of atoms of the side chains of residue $r$. Based on these new coordinates, we can update the matrix of node features, denoted by $\hat{\bm A}$, in which the features of other atoms in the graph are kept unchanged. The edges $\bm E$ of the complex during the disturbance are also unchanged.

\textbf{Reconstruction.} 
The pre-training scheme requires the GNN in GraphPPI to estimate the original coordinates of the given disturbed complex. However, as the ranges of the coordinates differ a lot in individual complexes, directly predicting the absolute values of the coordinates increases the difficulty of reconstruction. Instead, GraphPPI accomplishes the reconstruction by predicting the difference in coordinates of the atoms in the disturbed residue. 

More specifically, we first feed the initial atom features $\hat{\bm A}$ of the disturbed complex into the GNN and obtains the corresponding graphical descriptors $\hat{\bm G}$ for all the atoms.
\begin{align}
\hat{\bm G} =  \{\hat{\bm g}_k|k=1,2,\dots,N\} = \text{CGAT}(\hat{\bm A}, \bm E).
\end{align}
Based on the graphical descriptors $\hat{\bm g}_k$ of node $k$ generated by the GNN, GraphPPI employs a multi-layer perceptron network (MLP) to predict the changes of the coordinate, that is,
\begin{align}
 (\triangle {x}_k,\triangle {y}_k,\triangle {z}_k) &= \text{MLP}(\hat{\bm g}_k).
\end{align}
Thus, the predicted coordinate of node $k$ can be derived by
\begin{align}
 (\tilde{x}_k,\tilde{y}_k,\tilde{z}_k) &= (\hat{x}_k,\hat{y}_k,\hat{z}_k) +(\triangle {x}_k,\triangle {y}_k,\triangle {z}_k)
\end{align}

The reconstruction loss of GraphPPI is the mean square errors between the predicted coordinates with the original coordinates of the disturbed atoms, given by
 
\begin{align}
\mathcal{J} &= \frac{1}{|S(r)|}\sum_{k\in S(r)} [(x_k-\tilde{x}_k)^2+ (y_k-\tilde{y}_k)^2+(z_k-\tilde{z}_k)^2],
\end{align}
where $|S(r)|$ is the cardinality of the set $S(r)$.

\clearpage

\small
%dcu
\bibliographystyle{dcu}
\bibliography{main}
\clearpage

\newpage
\renewcommand{\figurename}{Supplementary Figure}
\renewcommand{\tablename}{Supplementary Table}

\setcounter{page}{1}
\setcounter{figure}{0}
\setcounter{table}{0}
\title{\bf{Supplementary Information for \\  Pre-training of Graph Neural Network for Modeling Protein-Protein Binding Affinity Changes Following Mutation}}
\maketitle

\symbolfootnote[0]{\hspace{-0.1in} \footnotesize{$^1$} Department of Computer Science, University of Illinois at Urbana
Champaign, IL, USA.}

\symbolfootnote[0]{\hspace{-0.1in} \footnotesize{$^2$} Laboratory for Brain and Intelligence and Department of Biomedical Engineering, Tsinghua University, Beijing,
China.}
\symbolfootnote[0]{\hspace{-0.1in} \footnotesize{$^3$} Beijing Innovation Center for Future Chip, Tsinghua University, Beijing, China.}

\section*{Supplementary Figures}

\begin{figure}[ht]
\centering
\includegraphics[width=0.6\linewidth]
{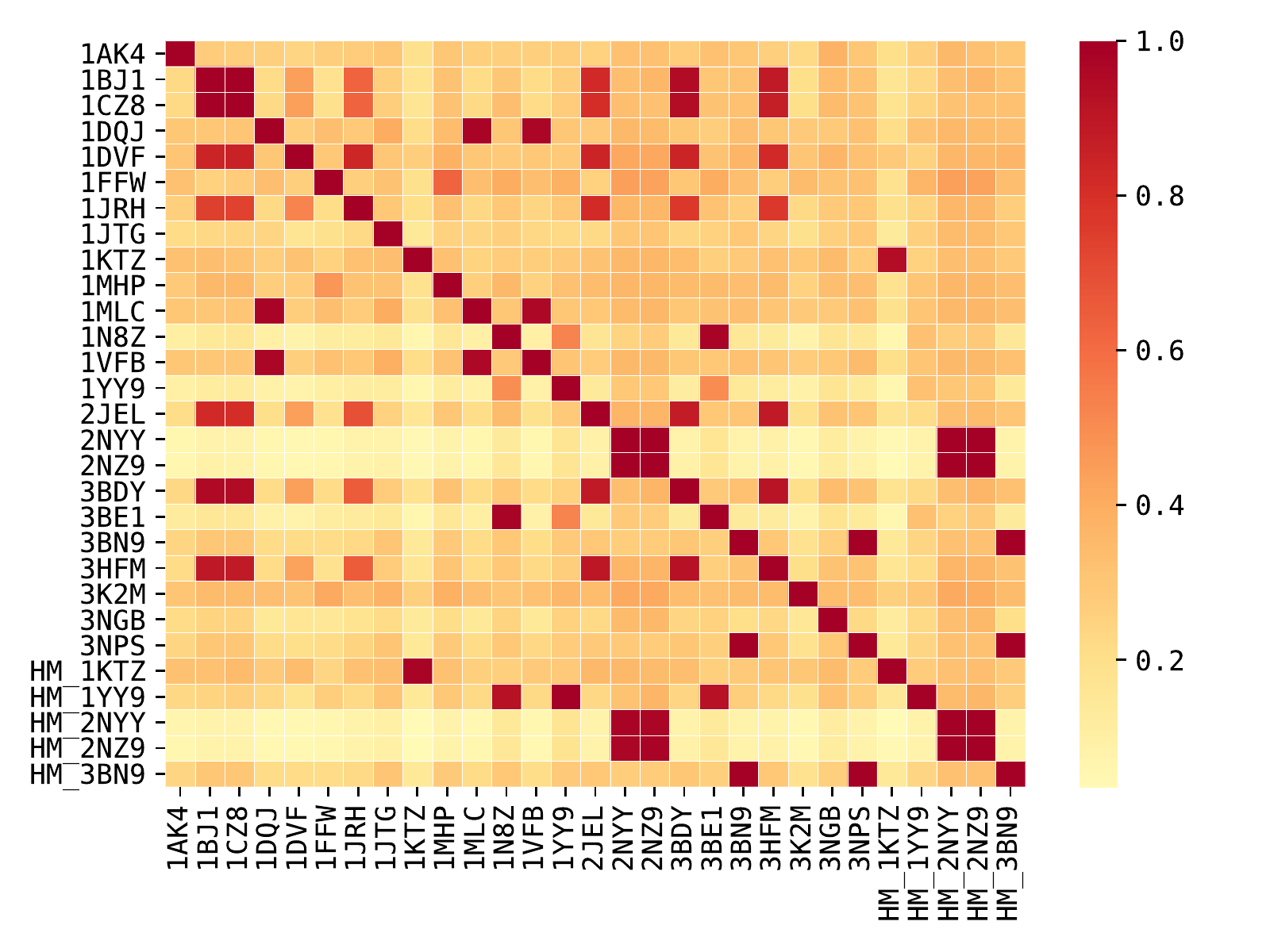}
\caption{The similarities between arbitrary two complexes in the s645 set.}
\label{fig:s645-sim}
\end{figure}

\begin{figure}[ht]
\centering
\includegraphics[width=0.8\linewidth]
{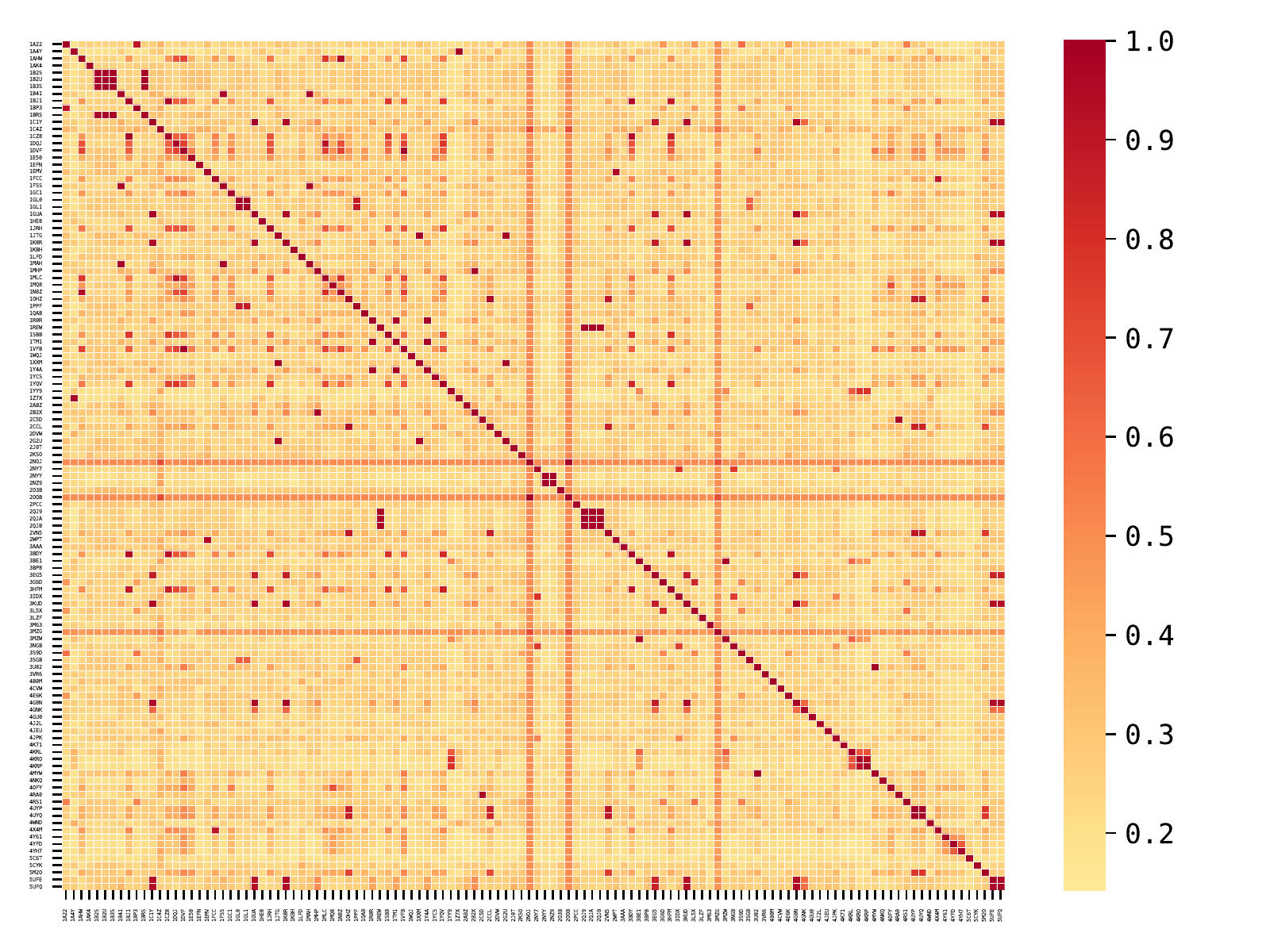}
\caption{The similarities between arbitrary two complexes in the M1707 set.}
\label{fig:m1707-sim}
\end{figure}

\clearpage
\begin{figure}[t]
\centering
\includegraphics[width=0.8\linewidth]
{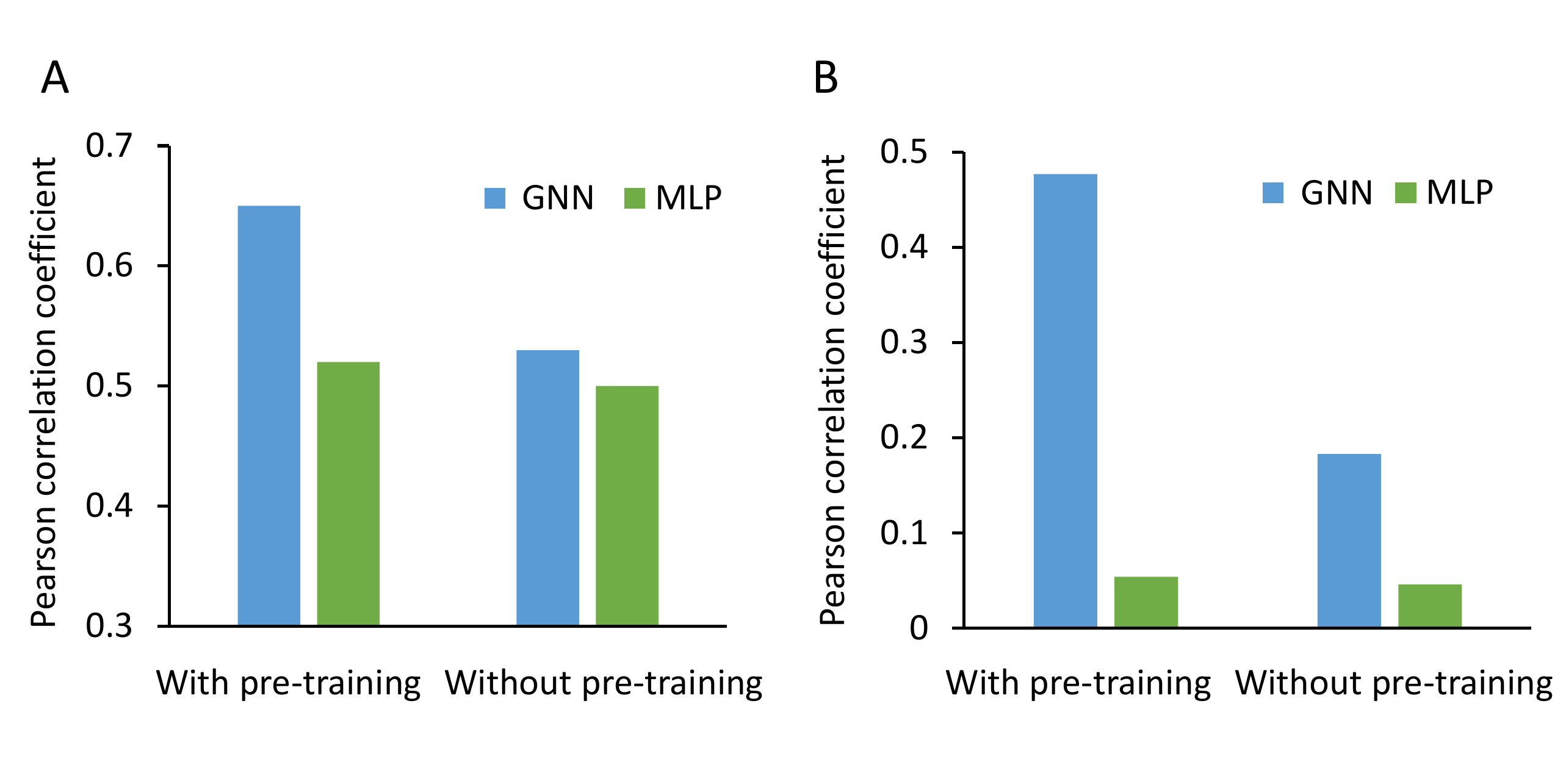}
\caption{The ablation Study on AB-Bind S645 set. (A) The prediction performance of GraphPPI with different pre-training strategies and transformation layers (i.e., CGAT, MLP) on the ten-fold CV test. (B) The prediction performance of GraphPPI with different pre-training strategies and transformation layer on the leave-out-one-complex CV test. To test the effectiveness of the GNN model. we built a control framework that uses a multiple layer perceptron (MLP) to replace the GNN. More specifically, each multi-attention transformation layer (Equation (\ref{eq:multi-att})) was replaced by an MLP layer. The main difference between the GNN and MLP lies in the way of processing the information of neighboring nodes. For a node in the graph, MLP updates the representations based on its own representations, while the GNN is also able to integrate the information from the neighboring nodes. 
}
\label{fig:ablation}
\end{figure}

\clearpage

\begin{table}[!htb]
	\caption{
	The optimal hyperparameters of GraphPPI calibrated using a grid search procedure. The hyperparameter tuning process involved the hidden size $D_g$, the number of attention heads $K$, number of hidden layers $L$.  We applied a coarse grid search approach over $D_g \in  \{128,256,512\}$, $K \in \{2,4,6,8,16\}$,  $L\in\{1,2,3,4,5\}$ on the development set to select the best settings of these hyperparameters. } 
	\begin{center}
		\begin{tabular}{ccccc}
			\hline\noalign{\smallskip}
			Hyperparameter & Selected values  \\
			\noalign{\smallskip}
			\hline
			\noalign{\smallskip}
			$D_g$  & 256     \\
			$K$     &  8     \\
			$L$        & 4   \\
			\hline
		\end{tabular}
	\end{center} 
	\label{table:hyper}
\end{table}

\begin{table}[ht]
	\center
     \caption{Node features and corresponding encoding methods.}
	\tiny
	\resizebox{0.6\linewidth}{!}{
		\begin{tabular}{ccc}
			\hline\noalign{\smallskip}
			\textbf{Features}  & Encoding method & Dimension \\		
			\noalign{\smallskip}
			 \hline		\noalign{\smallskip}
			Atom type (C,N,O,S) & One-hot encoding & 4\\ 
			\noalign{\smallskip}
			Residue type & One-hot encoding & 20\\
			\noalign{\smallskip}
			Is in the mutated chain & Binary values & 1 \\
			\noalign{\smallskip}
			Chain index & Integer & 1 \\ 
			\noalign{\smallskip}
			Is on the interface  &  Binary values& 1\\
			\noalign{\smallskip}
			Position  $({x},{y},{z})$ & Numeric values& 3 \\
			\hline
		\end{tabular}
	}

	\label{tab:features}
\end{table}

\end{document}